# Low-Frequency Noise Characteristics of GaN Vertical PIN Diodes – Effects of Current and Temperature

Subhajit Ghosh[1], Kai Fu[2], Fariborz Kargar[1], Sergey Rumyantsev[3], Yuji Zhao[2], and Alexander A. Balandin[1]*[1]

[1]Nano-Device Laboratory, Department of Electrical and Computer Engineering, Bourns College of Engineering, University of California, Riverside, California 92521 USA

[2]Department of Electrical and Computer Engineering, Rice University, Houston, TX 77005 USA

[3]CENTERA Laboratories, Institute of High-Pressure Physics, Polish Academy of Sciences, Warsaw 01-142 Poland

## Abstract

We report low-frequency noise characteristics of vertical GaN PIN diodes, focusing on the effects of the diode design, current and temperature. The as-grown and regrown diodes, with and without surface treatment have been studied. The noise in most of the GaN devices had a characteristic *1/f* spectrum at high and moderate currents, while some devices revealed generation-recombination bulges at low currents (*f* is the frequency). The predominant trend of the noise spectral density, $S_I$, dependence on the current was $S_I \sim I$. All tested GaN PIN diodes had rather low normalized noise spectral densities of $10^{-18}$ cm$^2$/Hz – $10^{-16}$ cm$^2$/Hz (*f*=10 Hz) at the current density *J*=1 A/cm$^2$ at room temperature. The noise temperature dependences at different currents revealed peaks at T=375 K – 400 K. Temperature, current, and frequency dependences of noise suggest that the noise mechanism is of the recombination origin. We argue that the noise measurements at low currents can be used to efficiently assess the quality of GaN PIN diodes.

**Keywords:** low-frequency noise; gallium nitride; PIN diode; temperature dependence

---

[1] Corresponding author; e-mail: balandin@ece.ucr.edu ; https://balandingroup.ucr.edu/

Development of the next generation of GaN PIN diodes for high-power electronics requires effective methods for assessing materials and device quality. Low-frequency noise measurements have been widely used for the characterization of defects in various semiconductor devices, and for testing their reliability [1-14]. The noise level and its current or gate voltage dependence in the field-effect transistors can be used to compare the quality of the device structures. Temperature dependence of the low-frequency noise is often used to determine its origin and physical mechanism. Accurate knowledge of the specific semiconductor devices' noise characteristics is also required for circuit level modeling. The information on the low-frequency noise characteristics of GaN PIN diodes is limited [15]. No direct comparison between different technologies is available. There are no detailed studies of how chemical treatment and etching, used in device fabrication, affect the noise level of GaN PIN diodes.

Selective-area doping is a key to the fabrication of high-performance advanced GaN power devices [16]. Epitaxial regrowth is the preferable method for GaN PIN diodes fabrication compared to diffusion and ion implantation for selective-area doping [16-19]. Some of us have previously reported selective *p*-doping by regrowth technologies that included dry etching before the regrowth [16-18]. Optimization of the etch-then-regrow process is important for further development of GaN high-power electronics. The regrowth often results in a high concentration of impurities, including silicon, carbon, and oxygen; the dry etching process can also introduce surface damage [16-18]. The impurity atoms and surface defects can act as trapping and recombination centers, potentially leading to a substantial increase in low-frequency noise. In this Letter, we report an investigation of noise characteristics of GaN PIN diodes and demonstrate that the low-frequency noise at low currents is a convenient and useful figure-of-merit for the quality of GaN PIN diodes.

The GaN layer were grown on c-plane n-GaN free-standing conducting substrates by metalorganic chemical vapor deposition (MOCVD). Several different technologies have been considered: the etch-then-regrow structures with relatively high interface quality obtained by the low RF power etching and UV-Ozone and chemical treatment (Type I); etch-then-regrow structures with lower interface quality obtained without the UV-Ozone or chemical treatment (Type II); as-grown devices with the guard rings (Type III). In Type I devices, the low-power RF dry etching along with UV-Ozone and chemical treatments were used to form high-quality regrowth interfaces. The Type II devices have the same regrown structure as Type I but were fabricated using a higher etching RF power (70 W) without the surface treatment. The Type III devices were based on as-grown structures without a regrowth interface. A guard ring structure was made on Type III devices to improve the breakdown voltage [19]. Metal contacts were deposited using electron-beam evaporation to form the anode and cathode contacts for all the devices. Additional details of the fabrication process, including the geometry of the etched regions, chemical treatment recipes, have been reported elsewhere [16-19]. In order to better correlate the noise level with the leakage current, we selected an additional regrown device, fabricated without etching or chemical treatment, which had relatively large leakage current. We refer to this reference device as Type IV. The layered structures for studied devices are shown in Figure 1.

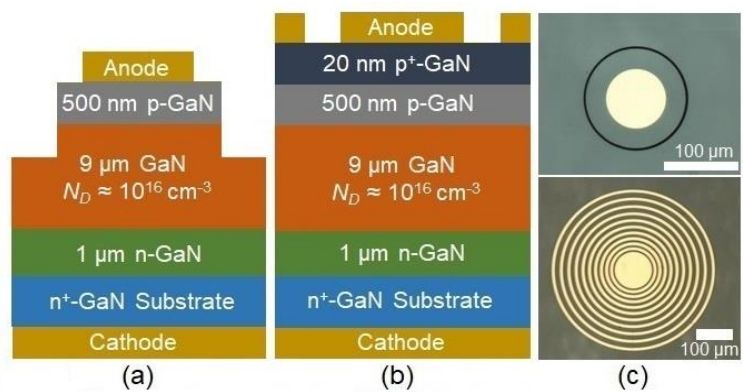

**Figure 1:** Schematics of the GaN PIN diode structures. (a) The layered structure of the regrown devices of Type I, II, and IV. Type I is the regrown devices with the low RF power dry etching and UV chemical treatments; Type II is the regrown devices with the high RF power dry etching without any chemical treatment; Type IV is the regrown device without chemical treatment with a relatively large leakage current, intentionally selected as a reference structure. (b) The layered structure of the Type III as-grown devices with guard rings. (c) Optical microscopy images of a regrown Type II device with a mesa (top) and as-grown Type III device (bottom).

The current-voltage (I-V) and low-frequency noise characteristics for all types of the PIN diodes were measured at room temperature in vacuum (Agilent B1500; Lake Shore TTPX). The noise spectra were acquired with a dynamic signal analyzer (Stanford Research 785). The signal analyzer was used to measure the absolute voltage-referred noise spectral density, $S_V$, on a load resistor, $R_L$, in series with the device under test, $R_D$. The load resistor was grounded in our configuration. A potentiometer was used to control the voltage supplied with a low noise battery. During the noise measurements, the voltage fluctuations were transferred to a low-noise preamplifier; then the amplified time domain signal was transformed to its corresponding frequency domain using a dynamic signal analyzer. The spectral density, $S_V$, was recalculated to the current spectral density $S_I$, and then normalized by the current squared, $I^2$, and the cross-section area of the PIN diodes. Details of the noise measurements procedures, in the context of other material systems, have been reported by some of us elsewhere [20-23].

We start by comparing I-V and noise characteristics of different PIN diodes belonging to the same type. Figure 2 (a) shows I-Vs for three representative devices of Type I. At the forward voltage, $V_F$, in the range 3 V>$V_F$>2.5 V the I-V characteristics are exponential with the ideality factor ~2 for all devices. At $V_F$ >3 V, the characteristics in the semi-logarithmic scale tend to saturate indicating the dominant contribution of the series resistance. At low bias $V_F$ <2.5 V, the devices of the same type behave differently demonstrating the different levels of the leakage current. The leakage current is the highest for device A, followed by devices B, and C. Among these three tested diodes, device C demonstrated the best performance. Figure 2 (b) shows the bias points for the noise measurements in the device A. The device with the higher leakage currents was selected intentionally to illustrate the correlation between I-Vs and noise characteristics. In Figures 2 (c), we present the current noise spectral density, $S_I$, for device A as a function of frequency, $f$, at different currents. One can see that the noise is of the general $1/f$ type, particularly at higher currents, *i.e.*, at $I$= 0.01 A and 0.005 A. At lower currents, $I \leqslant$ 50 μA, the noise spectra reveal the generation – recombination (G-R) bulges. From analysis of the data, we concluded that devices with larger leakage currents are typically those that have G-R features in their noise spectra (see Supplemental Materials for more data). The noise in the device C, with the lowest leakage current, was of the $1/f$ type. These data suggest that G-R noise is caused by the defects acting as the recombination centers, which are responsible for the higher leakage currents at low bias.

It is interesting that the noise at the intermediate current of $I=5\times10^{-5}$ A is the smallest, *i.e.*, G-R noise depends on current non-monotonically, and it has a minimum at some current. Similar behavior was reported previously for SiC *p-n* junction [24]. The model of the G-R noise developed in Ref. [24] confirms this kind of non-monotonical dependence. Therefore, we can conclude that the high leakage current in this device is not due to some parasitic channel leakage but rather due to the recombination in the space charge region. The current and area normalized noise spectral densities versus current density, at a fixed frequency $f=10$ Hz, are presented in Figure 2 (d) for all three devices. One can see that in all devices, the normalized noise spectral density decreases with the current density. The noise characteristics at low current density, i.e., $J$<1 A/cm$^2$, correlate perfectly the GaN PIN diode performance, i.e., the devices with the lowest leakage current have the lowest noise level. The low-frequency noise and the leakage current, are defined by the nature and concentration of the defects introduced during the processing steps which act as recombination centers. Therefore, the noise measured at low bias can be a sensitive metric of the device quality.

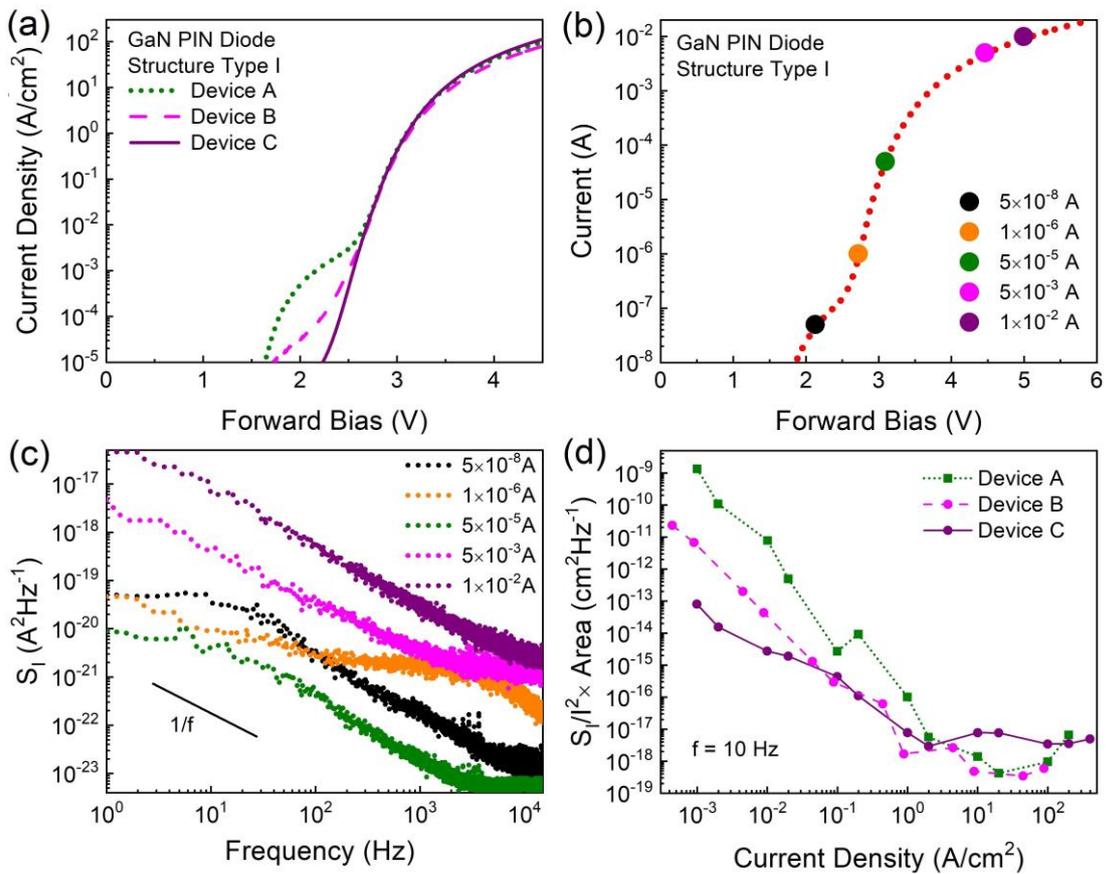

**Figure 2**: Electrical and noise characteristics of three different devices of the same Type I. (a) Current-voltage characteristics of the Type I devices, A, B, and C, with different leakage currents. (b) Current-voltage characteristics of device A with colored dots indicating the noise measurement data points. (c) Low-frequency noise spectra for different currents indicated in panel (b). Note G-R bulges in the noise spectra at low currents. At high currents, the noise is of the typical *1/f* type. (d) The normalized current noise spectral density, $S_I/I^2$, as a function of the current density for three devices measured at $f=10$ Hz.

Figure 3 (a) shows forward I-V characteristics of all four types of GaN PIN diodes. We selected device C from the three previously measured Type I devices owing to its performance and low noise level. In Figure 3 (b) we present the I-V of a representative Type III as-grown device that had the best rectifying performance. Note that this I-V characteristic is close to a classical one. At low currents, the ideality factor $n_1$=2.1, which is close to the ideal recombination current, as it should be at low bias. At the current density $J$>10$^{-2}$ A, the diffusion current starts to contribute, leading to the ideality factor decrease to $n_2$=1.6. Figure 3 (c) shows the current noise spectral density as a function of the current density at a fixed frequency $f$=10 Hz for all types of the devices. The data are plotted for each type of GaN PIN diodes with the corresponding I-Vs shown in Figure 3 (a). There are several important observations. The $S_I$ vs. $J$ ($S_I$ vs. $I$) relation follows the trend $S_I \sim J$ ($S_I \sim I$), observed in many different $p$-$n$ and Schottky diodes. At low currents, one can notice the deviation from this trend, particularly for the Type II and Type IV devices. We attribute this deviation to the appearance of the G-R bulges in these devices at low currents. This is in line with the explanation of a higher concentration of some specific defects, acting as recombination centers, which result in higher leakage current and noise level. One should note that noise with 1/$f$ spectrum can also be a superposition of G-R noise bulges from several distinctive defects which are characterized by different time constants [25]. A prior study of GaN/AlGaN PIN diodes of a different design on a sapphire substrate reported $S_I \sim I^2$, similar to a linear resistor [26]. However, there have been reports for AlGaN PIN photodetector diodes with $S_I \sim I$ [27]. It is interesting to note that noise in the studied diodes is at least four to seven orders of magnitude lower than that reported for GaN/AlGaN PIN diodes in Ref. [15] and comparable to those in the lateral GaN/AlGaN Schottky diodes [28].

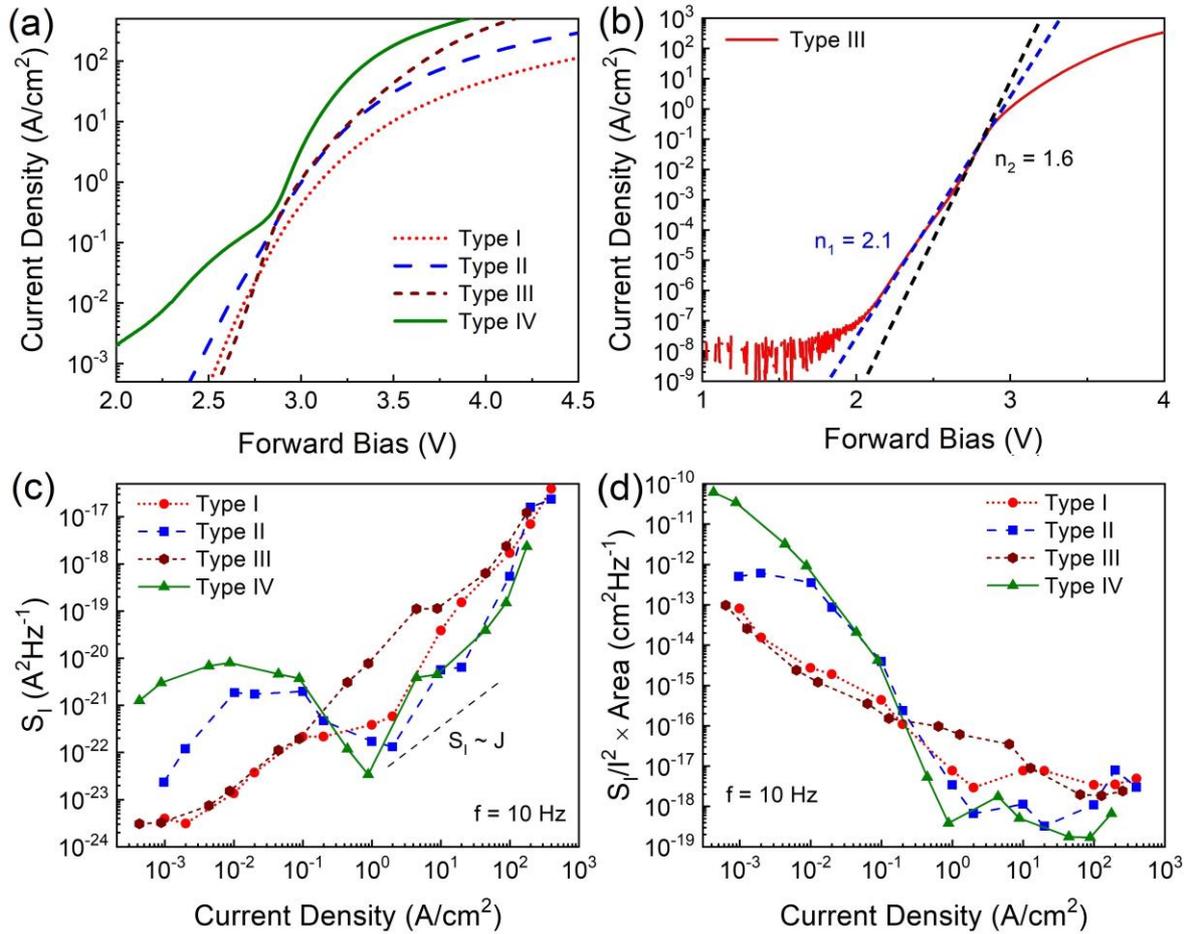

**Figure 3**: Electrical and noise characteristics of GaN PIN diodes of different type. (a) Current-voltage characteristics of representative devices of each type. (b) The forward bias current density for Type III as-grown PIN diode clearly showing the ideality factor. Note that the ideality factor takes values of 2.1 and 1.6 in the recombination and diffusion regions, correspondingly. (c) The current noise spectral density, $S_I$, as a function of the current density, $J$, at $f$ = 10 Hz. (d) The normalized current noise spectral density, $S_I/I^2$, as a function of the current density at $f$ = 10 Hz. The normalized noise level is correlated with the respective electrical properties at low currents.

For a more accurate and direct comparison of the noise level in each GaN PIN technology, in Figure 3 (d), we plot the current noise spectral density normalized by the current squared and device area, $S_I/I^2 \times \Omega$, vs. current density, $J$, at fixed $f$=10 Hz ($\Omega$ is the area of the top contact). One can see that at the small currents, $J$< 0.1 A/cm$^2$, the lowest noise is in the diodes of Type

7 | P a g e

I and Type III while the highest level of noise is in the reference Type IV PIN diode which is characterized by the high leakage current. This observation suggests that the chemical treatment and etching in our devices did not result in the strongly increased noise. We also note that the noise characteristics at low currents are the most informative since they are defined by the carrier recombination with participating of slow processes that contribute to the low-frequency noise. For all devices, the normalized noise spectral density decreases with the increasing current density. At higher currents, the noise $S_I/I^2$ tends to be independent on the current, which is typical behavior for the noise from a linear resistor. This noise is due to the series resistance of the devices.

To shed light on the physical mechanism of noise in these devices, we measured noise as a function of temperature. Measurements at elevated, rather than low temperatures, have proven to be informative for devices based on wide-band-gap semiconductors [29-32]. The elevated temperatures are also more relevant to the intended high-power applications of GaN PIN diodes. In Figure 4, we present the temperature dependent electrical (Figure 4 (a)) and noise characteristics (Figure 4 (b-d)) of a representative as-grown GaN PIN diode, defined above as Type III. The I-V characteristics are exponential at all studied temperatures with the ideality factor only weakly dependent on temperature. It means that even at elevated temperatures and at low bias the current is still of recombination origin without significant contribution of the parasitic leakage. It is interesting to note that the noise spectral density dependences on the current density reveal peaks for the temperatures in the range from $T=375$ K to $T=450$ K. Three regions with the changing as a function of temperature, $S_I \sim I^\zeta$, dependence can be distinguished in Figure 4 (c). In region I, $\zeta$ changes approximately from ~1 to ~2; in region II, it varies between ~1 and some negative values; and in the high-current region III, $\zeta$ stabilizes to around ~2, as one observes in linear resistors (see Supplemental Materials). The noise dependence on temperature (Figure 4 (d)) reveals peaks for the current levels that correspond to the transition from region I to II (in Figure 4 (c)). The current dependences of noise at elevated temperatures resemble those observed for the devices with the elevated leakage current and G-R noise spectra (Figure 2). Temperature and current dependences of noise, which demonstrate maxima, are characteristic features of G-R origin of noise [24]. The latter suggest that the 1/$f$ noise in the devices of Type III is still of the G-R origin. The increase of the recombination current at elevated temperatures helps to reveal this noise mechanism. Since the noise spectra shapes are still 1/*f*-like we have to conclude that at least several levels with different characteristic times contribute to noise within the studied frequency band.

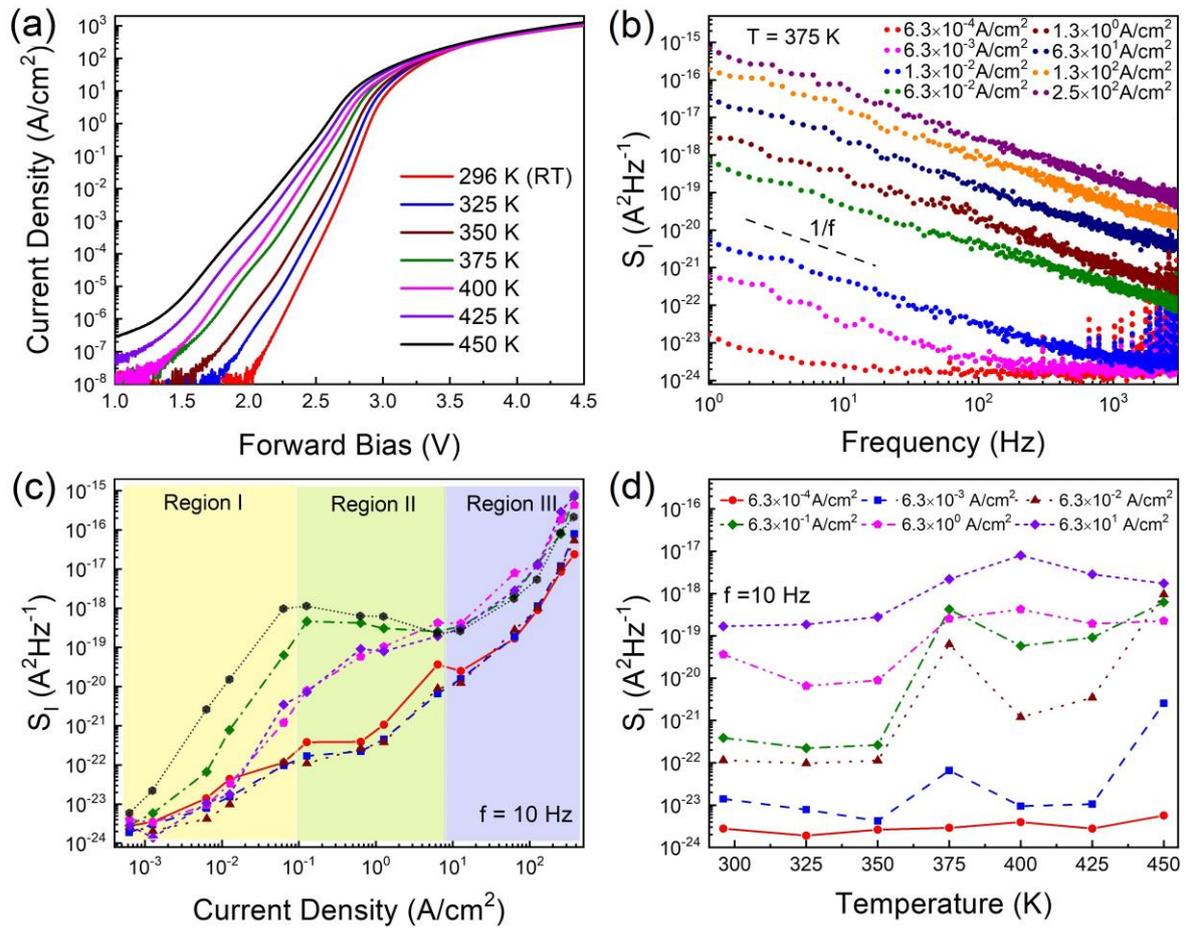

**Figure 4**: Temperature dependent electrical and noise characteristics of GaN PIN diode of Type III. (a) Current-voltage characteristics of a representative GaN PIN diode at elevated temperatures. (b) The current noise spectral density, $S_I$, as a function of frequency measured at different currents. (c) The current noise spectral density, $S_I$, as a function of the forward current density at $f = 10$ Hz shown for different temperatures. The colors of the curves correspond to the temperatures indicated in the (a) panel. (d) The current noise spectral density, $S_I$, as a function of the temperature $T$, shown for different current densities at $f = 10$ Hz.

In conclusion, we reported on a study of low-frequency noise in vertical GaN PIN diodes of different designs and process technologies. The noise characteristics were measured at different current densities and elevated temperatures relevant to the high-power switch applications. The noise in all considered device types has a characteristic *1/f* spectrum at high currents. Some devices with the largest leakage current revealed G-R bulges at low currents. The noise spectral density, $S_I$, predominantly scales with the current, $I$, as $S_I \sim I$. Out data indicate that the noise measurements at low currents can be used to efficiently assess the quality of GaN PIN diodes.

**Acknowledgements**

The work at UC Riverside and Rice University was supported by ULTRA, an Energy Frontier Research Center (EFRC) funded by the U.S. Department of Energy, Office of Science, Basic Energy Sciences under Award # DE-SC0021230. S.R. who contributed to noise data analysis acknowledges the support by CENTERA Laboratories in a frame of the International Research Agendas program for the Foundation for Polish Sciences co-financed by the European Union under the European Regional Development Fund (No. MAB/2018/9).

**Author Contributions**

A.A.B. coordinated the project, lead data analysis and manuscript preparation. S.G. measured current-voltage and low-frequency noise characteristics; K.F. and Y.Z. fabricated GaN devices and conducted current-voltage measurements; S.R. and F.K. contributed to the noise data analysis. All authors contributed to the manuscript preparation.

**Supplemental Information**

The supplemental information is available at the Applied Physics Letters journal web-site for free of charge.

**The Data Availability Statement**
The data that support the findings of this study are available from the corresponding author upon reasonable request.

**REFERENCES**

1. A. A. Balandin, *Noise and Fluctuations Control* (American Scientific Publishers, 2002).
2. L.K.J. Vandamme, IEEE Transactions on Electron Devices **41**, 2176 (1994).
3. P. Dutta and P.M. Horn, Reviews of Modern Physics **53**, 497 (1981).
4. A. Balandin, S. Morozov, G. Wijeratne, S. J. Cai, R. Li, J. Li, K. L. Wang, C. R. Viswanathan, and Y. Dubrovskii, Applied Physics Letters, **75**, 13 (1999).
5. S.L. Rumyantsev, N. Pala, M.S. Shur, E. Borovitskaya, A.P. Dmitriev, M.E. Levinshtein, R. Gaska, M.A. Khan, J. Yang, X. Hu, and G. Simin, IEEE Transactions on Electron Devices **48**, 530 (2001).
6. J. Tartarin, G. Soubercaze-Pun, J. Grondin, L. Bary, J. Mimila-Arroyo, and J. Chevallier, AIP Conference Proceedings **922**, 163 (2007).
7. A. V. Vertiatchikh and L.F. Eastman, IEEE Electron Device Letters **24**, 535 (2003).
8. T.M. Chen and A.M. Yassine, IEEE Transactions on Electron Devices **41**, 2165 (1994).
9. H. Rao and G. Bosman, Journal of Applied Physics **108**, 053707 (2010).
10. H.P. Rao and G. Bosman, IEEE Transactions on Device and Materials Reliability **12**, 31 (2012).
11. M. Silvestri, M.J. Uren, N. Killat, D. Marcon, and M. Kuball, Applied Physics Letters **103**, 043506 (2013).
12. D.M. Fleetwood, S. Beyne, R. Jiang, S.E. Zhao, P. Wang, S. Bonaldo, M.W. McCurdy, Zs. Tőkei, I. DeWolf, K. Croes, E.X. Zhang, M.L. Alles, R.D. Schrimpf, R.A. Reed, and D. Linten, Applied Physics Letters **114**, 203501 (2019).
13. N.A. Hastas, C.A. Dimitriadis, and G. Kamarinos, Applied Physics Letters **85**, 311 (2004).
14. I.-H. Lee, A.Y. Polyakov, S.-M. Hwang, N.M. Shmidt, E.I. Shabunina, N.A. Talnishnih, N.B. Smirnov, I. v. Shchemerov, R.A. Zinovyev, S.A. Tarelkin, and S.J. Pearton, Applied Physics Letters **111**, 062103 (2017).
15. L. Dobrzanski and W. Strupinski, IEEE Journal of Quantum Electronics **43**, 188 (2007).
16. K. Fu, H. Fu, H. Liu, S.R. Alugubelli, T.-H. Yang, X. Huang, H. Chen, I. Baranowski, J. Montes, F.A. Ponce, and Y. Zhao, Applied Physics Letters **113**, 233502 (2018).
17. K. Fu, H. Fu, X. Huang, H. Chen, T.H. Yang, J. Montes, C. Yang, J. Zhou, and Y. Zhao, IEEE Electron Device Letters **40**, 1728 (2019).
18. K. Fu, H. Fu, X. Deng, P.-Y. Su, H. Liu, K. Hatch, C.-Y. Cheng, D. Messina, R.V. Meidanshahi, P. Peri, C. Yang, T.-H. Yang, J. Montes, J. Zhou, X. Qi, S.M. Goodnick, F.A. Ponce, D.J. Smith, R. Nemanich, and Y. Zhao, Applied Physics Letters **118**, 222104 (2021).
19. H. Fu, J. Montes, X. Deng, X. Qi, S.M. Goodnick, F.A. Ponce, Y. Zhao, K. Fu, S.R. Alugubelli, C.Y. Cheng, X. Huang, H. Chen, T.H. Yang, C. Yang, and J. Zhou, IEEE Electron Device Letters **41**, 127 (2020).
20. A.A. Balandin, Nature Nanotechnology **8**, 549 (2013).
21. G. Liu, S. Rumyantsev, M.A. Bloodgood, T.T. Salguero, and A.A. Balandin, Nano Letters **18**, 3630 (2018).
22. A. Geremew, C. Qian, A. Abelson, S. Rumyantsev, F. Kargar, M. Law, and A. A. Balandin, Nanoscale **11**, 20171 (2019).

23. S. Ghosh, F. Kargar, A. Mohammadzadeh, S. Rumyantsev, and A.A. Balandin, Advanced Electronic Materials, 2100408 (2021).
24. S.L. Rumyantsev, A.P. Dmitriev, M.E. Levinshtein, D. Veksler, M.S. Shur, J.W. Palmour, M.K. Das, and B.A. Hull, Journal of Applied Physics **100**, 064505 (2006).
25. M. V. Haartman, M. Östling, *Low-Frequency Noise in Advanced MOS Devices* (Springer, 2007).
26. L.S. Ma, Z. Bi, A. Bartels, K. Kim, L. Robertsson, M. Zucco, R.S. Windeler, G. Wilpers, C. Oates, L. Hollberg, and S.A. Diddams, IEEE Journal of Quantum Electronics **43**, 139 (2007).
27. T. Li, D.J.H. Lambert, M.M. Wong, C.J. Collins, B. Yang, A.L. Beck, U. Chowdhury, R.D. Dupuis, and J.C. Campbell, IEEE Journal of Quantum Electronics **37**, 538 (2001).
28. G. Cywiński, K. Szkudlarek, P. Kruszewski, I. Yahniuk, S. Yatsunenko, G. Muzioł, C. Skierbiszewski, W. Knap, and S.L. Rumyantsev, Applied Physics Letters **109**, 033502 (2016).
29. Y.-Y. Chen, Y. Liu, Y. Ren, Z.-H. Wu, L. Wang, B. Li, Y.-F. En, and Y.-Q. Chen, Modern Physics Letters B **35**, 2150134 (2020).
30. D. V. Kuksenkov, H. Temkin, A. Osinsky, R. Gaska, and M.A. Khan, Applied Physics Letters **72**, 1365 (1998).
31. D. V. Kuksenkov, H. Temkin, A. Osinsky, R. Gaska, and M.A. Khan, Journal of Applied Physics **83**, 2142 (1998).
32. B.H. Leung, N.H. Chan, W.K. Fong, C.F. Zhu, H.F. Lui, C.K. Ng, K.C. Wong, and C. Surya, Proceedings of the IEEE Hong Kong Electron Devices Meeting **148** (2001).